Larousse Khosravi Khorashad and Christos Argyropoulos*

Department of Electrical and Computer Engineering, University of Nebraska-Lincoln, Lincoln, NE, 68588, USA
*E-mail: christos.argyropoulos@unl.edu

# Unraveling the temperature dynamics and hot electron generation in tunable gap-plasmon metasurface absorbers

**Abstract:** Localized plasmons formed in ultrathin metallic nanogaps can lead to robust absorption of incident light. Plasmonic metasurfaces based on this effect can efficiently generate energetic charge carriers, also known as hot electrons, owing to their ability to squeeze and enhance electromagnetic fields in confined subwavelength spaces. However, it is very challenging to accurately identify and quantify the dynamics of hot carriers, mainly due to their ultrafast time decay. Their non-equilibrium temperature response is one of the key factors missing to understand the short time decay and overall transient tunable absorption performance of gap-plasmon metasurfaces. Here, we systematically study the temperature dynamics of hot electrons and their transition into thermal carriers at various timescales from femto to nanoseconds by using the two-temperature model. Additionally, the hot electron temperature and generation rate threshold values are investigated by using a hydrodynamic nonlocal model approach that is more accurate when ultrathin gaps are considered. The derived temperature dependent material properties are used to study the ultrafast transient nonlinear modification in the absorption spectrum before plasmon-induced lattice heating is established leading to efficient tunable nanophotonic absorber designs. We also examine the damage threshold of these plasmonic absorbers under various pulsed laser illuminations, an important quantity to derive the ultimate input intensity limits that can be used in various emerging nonlinear optics and other tunable nanophotonic applications. The presented results elucidate the role of hot electrons in the response of gap-plasmon metasurface absorbers which can be used to design more efficient photocatalysis, photovoltaics, and photodetection devices.

# 1    Introduction

Localized plasmons formed in ultrathin metallic nanogaps can lead to robust plasmonic metasurface absorber designs,[1] where the incident electromagnetic radiation is almost fully absorbed at the resonance. The structure of a typical gap-plasmon absorber is composed of periodic metallic nanostripes or nanocubes separated from a metallic substrate by an ultrathin dielectric spacer layer.[1] Recently, perfect narrowband absorbers were able to efficiently absorb and, subsequently, detect linear or circularly polarized light by using nanocubes[2] or chiral nanostripes,[3] respectively. Several alternative innovative metamaterials have been proposed to achieve broadband or narrowband absorption of light by using different materials and subwavelength unit cell geometries.[4–10]

One of the most vital applications of perfect light absorbers is the generation of energetic hot electrons that can be utilized in photochemical catalysis,[11–15] photo-electron detectors,[3,16] solar cells,[17,18] and plasmonic nonlinear devices.[19–22] The effect of the generated hot electrons in the ultrafast transient absorption response was recently indirectly measured in periodic arrays of gold nanodisks or silver nanocubes over metallic substrates, where hot electrons were mainly generated in the formed nanogaps due to the enhanced field.[23,24] In particular, it was shown that the time resolved spectral reflection under ultrafast pulse illumination exhibits a rapid change with femtosecond timescale duration followed by a steady state response achieved in slower timescales on the order of picoseconds. These complicated pump-probe ultrafast time-domain reflection measurements provided an indirect detection of the non-equilibrium hot electron generation dynamics and can lead to ultrafast time-modulated plasmonic nanostructures.[25–27] However, it is very challenging to directly experimentally observe or accurately theoretically model the dynamic behavior of hot electrons in these configurations mainly due to their ultrafast dynamic response accompanied by extremely fast decay time. These challenges become even more pronounced in the case of a few nanometer gaps, where nonlocal effects are also required to precisely predict the induced electric field distribution.

In general, when a light beam interacts with metallic (plasmonic) structures, their free electron gas rapidly oscillates because of the polarization induced by the external electromagnetic wave. This non-thermal distribution of electrons is generated as a result of the coherent plasmon resonance decay.[13,28–31] What is dictated by the current scientific knowledge is that at this point low energy electrons at the ground state of the metal are suddenly stimulated to excited states leaving behind hot holes.[13] Next, the high non-thermal hot electrons are redistributed via mainly electron-electron scattering on hundreds of femtoseconds to a few picoseconds timescales. At this point, a tail of high energy hot carriers is formed below and above the Fermi level of the metal. The resulted non-equilibrium state of hot electrons have high enough energy that can be incorporated into the conduction band or overcome the Schottky barrier limit if the metal is placed adjacent to a semiconductor.[32–35] These hot electrons can be used in photoelectron driven chemical reactions or be applied as electron catalyzers.[36–39] The final step in the relaxation process of the non-equilibrium hot carriers is the slower energy decay via lattice heat dissipation happening on hundreds of picoseconds to few nanoseconds. At this point, the energized electrons reach an equilibrium state inside the metallic nanostructure leading to a steady state response in terms of their temperature distribution. Note that the electron temperature can reach very high values on the order of thousands of degrees Kelvin during the hot carrier generation process and can modify the metallic material properties in the ultrafast picosecond time scale.[40] The hot electrons then couple to slower phonons that cause the much lower lattice temperature of the metallic nanostructure achieved in the steady state.[13] The electron temperature dynamics depend on various material properties, such as the electron-lattice coupling factor, thermal conductivity, and heat capacity, as well as to the enhanced and usually non-uniform electric field distribution generated in the corresponding plasmonic structure.[23,41]

Although the dynamics of hot electron relaxation processes have been extensively studied for extended metallic surfaces,[13,41–43] it is still a challenge to understand these processes in plasmonic nanoscale systems where strong light-matter interactions combined with quantum effects at the nanoscale are dominant.[1] In this work, we examine the hot electron generation and temperature dynamics in a widely used absorber configuration known as gap-plasmon metasurface that can achieve exceptionally high local density of optical states.[1] We demonstrate rapidly enhanced hot electron generation rates due to the boosted electric field inside the nanogap of these nanostructures, an interesting effect without a direct counterpart in bulk plasmonic materials.[44] The two-temperature model (TTM) is used mainly to capture the transition time of hot non-thermal carriers into the thermal carriers and to study these nanophotonic absorbers under various pulsed laser illuminations with the goal to adequately compute the hot electrons non-equilibrium temperature dynamics. We demonstrate that the TTM can sufficiently describe the transition of electrons from a very high non-thermal temperature distribution to non-equilibrium energetic carriers and finally thermal carriers when equilibrium at the steady state is reached. The presented computations also include the nonlocal material description of metals based on the hydrodynamic model,[19,45,46] which is more accurate to simulate extremely nanoscale gaps. This model can precisely investigate the ultimate limit of electric field enhancement and, consequently, the maximum temperature change dynamics and hot electron generation rate by the presented nanostructures. Interestingly, the metallic material properties change in the ultrafast time scale regime mainly due to the elevated electron temperature. Based on this observation, we predict that the absorption spectrum is also affected leading to ultrafast tunable nanophotonic absorber designs. In addition, we characterize the hot to thermal transition performance of electrons in the studied plasmonic absorbers for different pulse duration laser illuminations. The obtained temperature dynamic results are also applied to quantify the ultimate damage threshold limit of the presented plasmonic metasurfaces. This result is useful in deriving the input power limits used in nonlinear optics, such as second and third harmonic generation, and tunable nanophotonic applications where it is essential to know the maximum input power sustained by the materials before the structure is permanently damaged. The maximum power prediction is determined by considering the induced lattice temperature under various pulsed illuminations before reaching the melting temperature of the used metallic materials. Our results unravel the physics of hot electron generation and temperature dynamics in gap-plasmon absorbers leading to tunable nanophotonic designs. They can be used to maximize their nonlinear reconfigurable performance along with a plethora of other emerging applications, such as efficient photodetection, electrochemistry, photocatalysis, and energy harvesting.[43]

## 2    Theoretical Model

The effect of non-equilibrium temperature dynamics of electrons and lattice is required to be considered to accurately model the tunable photothermal phenomena induced by the interaction of ultrashort laser pulses with plasmonic nanostructures. In our analysis, we use the TTM to correlate the ultrafast dynamic evolution of energized hot carriers in plasmonic systems with their induced temperature variations. The TTM has already shown its capability to compute laser ablation in bulky metallic materials due to ultrashort femtosecond pulses.[41,47] The same theoretical model has been recently utilized to study the hot-carrier relaxation dynamics in elongated plasmonic thin films via pump-probe optical spectroscopy[42] and ultrafast thermomodulation microscopy.[48] However, it has not been used before in the analysis of plasmonic nanocavities, similar to the current absorbers, where the electric field is substantially enhanced in the nanoscale as it will be shown later.

Normally, dielectric materials have close to zero electrical conductivity and do not play a significant role in the TTM equations, since the hot electron cloud is mostly generated in metals that have a high imaginary part of permittivity leading to thermal dissipation. In its simplest form, the TTM is composed of two partial differential equations shown by Equations 1 and 2 to describe the spatiotemporal changes of induced electron ($T_e$) and lattice ($T_L$) temperatures:[41,47]

$$C_e \frac{\partial T_e}{\partial t} = \vec{\nabla} \cdot (k_e \vec{\nabla} T_e) - g_{e-L}(T_e - T_L) + S(\vec{r}, t), \tag{1}$$

$$C_L \frac{\partial T_L}{\partial t} = \vec{\nabla} \cdot (k_L \vec{\nabla} T_L) + g_{e-L}(T_e - T_L). \tag{2}$$

Equations 1 and 2 are coupled via the electron-lattice coupling factor $g_{e-L}$, which is non-zero only in the metallic parts of the presented metasurface. The dielectric parts (air and alumina in the current structure) have zero electron temperatures and their thermal response is defined by the regular lattice heat equation given by Equation 2 without the coupling factor term. By setting the index $i \equiv e$ or $L$, where $e$ is electron and $L$ depicts the lattice, $C_i$ and $k_i$ are the locally defined heat capacity and thermal conductivity, respectively, of the used materials. Note that the lattice heat capacity $C_L$ is given by the ratio of the specific heat capacity $C_p$ divided by the density of each material. Moreover, the electron-lattice coupling factor $g_{e-L}$ and electron heat capacity $C_e$ change with the induced electron temperature values and their plots in the case of silver are shown in Figure. S1(a)-(b). The values of all the other parameters used in Equations 1 and 2 are listed in Table S1. $S(\vec{r}, t)$ can be defined as any external heating source. In our current model, this is electromagnetic heating, as it is explained later in detail. The electron thermal conductivity varies with the induced electron and lattice temperatures and is given by the expression:[49]

$$k_e = k_L \frac{BT_e}{AT_e^2 + BT_L}, \tag{3}$$

where $A$ and $B$ are empirical parameters defined for silver in Table S2 and $k_L$ is the lattice thermal conductivity of silver given in Table S1. This formula was derived by molecular dynamics calculations and the contour plot of the electron thermal conductivity computed by Equation 3 is shown in Figure S1(c).

As mentioned earlier, the induced electron and lattice temperatures are locally ($\vec{r}$) and temporally ($t$) varying and the temperature distribution is given by the function: $T_i = T_i(\vec{r}, t)$. The $S(\vec{r}, t)$ component in Equation 1 is the induced heat source (with units of $W/m^3$) which is due to the interaction of the laser pulse with the plasmonic absorber. To be more precise, $S(\vec{r}, t)$ consists the resistive losses caused by the external laser field interacting with matter at the nanoscale. The incident laser illumination from free space is always defined as a harmonic oscillating electric field presented by the following Equation 4 with input intensity computed by Equation 5:

$$\vec{E}_0(\vec{r}, t) = E_0 \cdot Re\left[e^{-i\omega t + i\vec{k}\cdot\vec{r}}\right], \tag{4}$$

$$I_0 = \frac{1}{2} c \varepsilon_0 |E_0|^2. \tag{5}$$

The amplitude of the incident field in Equations 4 and 5 is $E_0$, $\omega$ and $\vec{k}$ are the angular frequency and wave vector of the input laser, respectively, $c$ is the speed of light, and $\varepsilon_0$ is the vacuum permittivity. Therefore, the average value of the induced heat source is defined as:

$$S(\vec{r}) = \langle \vec{j}(\vec{r}, t) \cdot \vec{E}(\vec{r}, t) \rangle_t = \frac{1}{2} Re\left[\vec{E}(\vec{r}) \cdot \vec{j}(\vec{r})\right], \tag{6}$$

where $\vec{j}(\vec{r}, t) = Re[\vec{j}(\vec{r}) \cdot e^{-i\omega t}]$ is the generated current density distribution and $\vec{E}(\vec{r}, t) = Re[\vec{E}(\vec{r}) \cdot e^{-i\omega t}]$ is the total electric field along the system caused by the incident field $\vec{E}_0(\vec{r}, t)$. The time averaged operation is shown in Equation 6 by the symbol $\langle ... \rangle_t$. In our frequency domain computations, Equation 6 is applied to TTM by using a temporal Gaussian profile to imitate the envelope of the pulsed laser illumination in a similar way to previous papers relevant to thermoplasmonics.[50,51] The employed temporal pulse has a normalized Gaussian profile: $\exp\left[-((t - 3\tau)/t)^2\right]$, where $\tau$ is the pulse duration and the pulse maximum amplitude is always obtained at $3\tau$. In the case of femtosecond illumination, the temporal part of the pulse is plotted in Figure S1(d). In the current work, the RF module of COMSOL Multiphysics is used for the frequency domain electromagnetic wave simulations. It is coupled to the heat transfer module to perform the electron and lattice temperature dynamic TTM-based calculations. The combination of the non-equilibrium temporal thermal response with the electromagnetic model is not trivial, since it requires to solve the partially differential equation system given by Equations 1 and 2 that substantially modifies the conventional thermal equations solved by COMSOL. More details about the complicated multiphysics modeling approach developed in this project are provided in the Supplementary Material. While the currently presented full-wave electromagnetic simulations are generally more accurate than the various semi-analytical methods existing in the literature [52]-[53], which usually require several approximations to be solved, there is a possibility to find similar semi-analytical solutions for the optical response of the current structure.

In conventional gap-plasmon metasurface simulations, the enhanced electric field in the dielectric nanogap formed between the metallic resonator and the metal backplane produces abruptly terminated charges on each metallic surface interface. However, the induced electric field enhancement is limited by the nonlocal properties of the metallic materials, an issue that is more pronounced as the gap becomes extremely thin.[19,45,46] In our work, both local and nonlocal models are investigated and compared when applied to the TTM and hot electron generation rate calculations. More specifically, we use the hydrodynamic model approach to compute the nonlocal effect.[45] In the nonlocal model, the surface charge density is not abrupt along the metallic surface but extents inside the metal to angstrom scale distances determined by the Thomas-Fermi screening length $\lambda_{TF} \propto \beta/\omega_p$, where the parameter $\beta$ is proportional to the Fermi velocity ($v_F$) and $\omega_p$ is the Drude plasma frequency. Hence, the electric current density inside the metal resulted by the induced electric field is modified by using the nonlocal model:[45,54]

$$\beta^2 \nabla(\nabla \cdot \vec{j}(\vec{r})) + (\omega^2 + i\gamma\omega)\vec{j}(\vec{r}) = i\omega\omega_p\varepsilon_0\vec{E}(\vec{r}), \tag{7}$$

where $\gamma$ is the damping parameter, $\omega_p$ is the plasma frequency, i is the imaginary unit, $\varepsilon_0$ is the vacuum permittivity, and ω is the frequency. In our calculations, silver is used for the metallic portions of the plasmonic absorber and the Drude model parameters used in Equation 7 are: $\omega_p = 2\pi \cdot 2175 [THz]$ and $\gamma = 2\pi \cdot 4.35 [THz]$.[55] We also choose $\beta = v_{F-Silver} = 1.39 \times 10^6 \, m/s$, similar to relevant previous works.[45,46] More details about the nonlocal simulations are provided in the Supplementary Material.

## 3 Results and Discussion

Figure 1 demonstrates the under-study gap-plasmon metasurface absorber and the inset depicts the dimensions and materials of a single unit cell. The absorber design is composed of a periodic array of silver nanostripes placed over an alumina spacer layer forming a dielectric nanogap. These plasmonic metasurface absorbers exhibit promising optical nonlinear responses, such as four wave mixing, spontaneous parametric down-conversion, and third harmonic generation,**[20–22,56]** among other useful applications[1] mainly due to the substantially enhanced electric field distribution in the nanogap. The absorber design is composed of a periodic array of silver nanostripes placed over an alumina spacer layer forming a dielectric nanogap. The opaque structure is terminated by an **80nm** thick silver substrate film in order to fully block transmission. The square nanostripe dimensions are **40nm × 40nm** and the period of the nanostructure is **250nm**. The structure is surrounded by air and excited by a normal incident linear polarized electromagnetic wave with electric field along the x-axis to achieve a resonating response. The curvature radius of the rounded nanostripe edges is chosen **2nm** to be consistent with relevant experimental designs.[21,22] The bulk permittivity of silver is taken from the literature[57] and is used in all electromagnetic wave simulations in the local model case. In general, plasmonic resonances and field distributions can be influenced by slight changes in the metasurface geometry.[23] We explored the field enhancement variation in the nanogap and absorptance resonance shift when using a sharper edge nanostripe, instead of the current more realistic rounded geometry, with results shown in Figure S3 in the Supplementary Materials.

The first step to demonstrate the strong absorption resonant response of this nanostructure is to compute its absorptance *A* spectra (derived by the reflectance *R* spectra simple by using: *A=1-R*) as a function of the nanogap thickness. The thick metal backplane acts as a mirror that backscatters the incident electromagnetic wave leading to negligible transmittance. Figure 2(a) shows the electromagnetic calculations of the absorptance spectra for increasing nanogap thickness when either local (solid lines) or nonlocal (dotted lines) metal models are used. Note that we have also calculated the absorption cross section (energy loss integration) within the metallic components of the nanostructure (not shown here) and observed the same response as the absorptance demonstrated in Figure 2(a). In addition, we have always used normal incidence plane wave excitation with electric field linearly polarized along the x-axis where the plasmonic resonance is excited. The other linear polarization will not couple to the system and will not generate field enhancement in the nanogap, i.e., the currently presented metasurface is polarization dependent. However, it can easily become polarization independent without altering its performance if the nanostripes are replaced by an array of three-dimensional plasmonic nanocubes that are polarization insensitive.[1] We have also demonstrated in Figure S4 that the normal incidence excitation generates the strongest field enhancement within the nanogap. It is noted that the resonant wavelength predicted by the nonlocal model is always blueshifted compared to the corresponding resonance in the local model. Furthermore, the perfect absorptive performance of our structure declines as the nanogap thickness is increased. This is because the mirror charges created at the plasmon resonance on the bottom of the silver stripe and on the top surface of the metal substrate are located further apart from each other. As a result, the critical coupling to the incident radiation is lost, leading to a less localized and enhanced electric field inside the nanogap.

Next, we compute the electron and lattice temperatures induced in metal by using the TTM, as described in the previous section. Our simulation results are shown in Figure 2(b) in the case of 2nm nanogap thickness and for an incident ultrafast laser with pulse duration $\tau = 130\ fs$ and intensity $I_0 = 3\ GW/cm^2$, where the local model is used. The temperature is calculated at point *P* depicted in the inset of Figure 2(b), where maximum field enhancement is obtained leading to the highest temperature values. The electron temperature ($T_e$) exhibits an abrupt peak at the maximum of the ultrashort Gaussian pulse and rapidly decays to merge with the lattice temperature ($T_l$) for longer timescales when steady state is reached. The maximum lattice temperature reaches approximately $315\,K$ at time $t = 6.64\,ps$. The obtained equilibrium between electron and lattice temperatures at steady state is the typical behavior that has been obtained before by using TTM in large-scale metallic structures.[41,58] The presented TTM calculations clearly demonstrate the rapid non-equilibrium hot electron transition to thermal carriers with time. To check the validity of our new TTM-based multiphysics simulations, we have performed similar modeling for much simpler metallic geometries without nanoscale features, such as bulk iron or 200nm thick flat gold films, and the results are shown in Figure S2. They are in perfect agreement with previous relevant works based on analytical approaches which, however, are different to the current full-wave multiphysics simulations that, in principle, can be applied to accurately model any complicated nanostructure.[41],[49]

Figure 2(c) demonstrates the amplitude of the computed electric field enhancement distribution derived from the ratio of the total field induced by the incident wave over the incident field: $E_{enh} = |\vec{E}(\vec{r})/\vec{E}_0(\vec{r})|$. The result of this simulation corresponds to the $2nm$ gap geometry plotted at each resonance of the local (top panel) and nonlocal (bottom panel) models. It is directly proven by these results that the field enhancement is limited in the nonlocal model due to the screening effect, where the surface charges extent into the silver nanostripe, as clearly shown in the bottom panel of Figure 2(d) that depicts a magnified region close to the nanostripe corner. The electric field (and consequently the surface charges) penetrate the silver nanostripe only in the nonlocal model case, while the field is abruptly blocked by the metal when the local model is used, as demonstrated in the upper panel of Figure 2(d). The results derived by using the hydrodynamic nonlocal model are expected to coincide with the local model for sufficient large gap thicknesses.[45] To demonstrate this point, the absorption resonance wavelengths and corresponding maximum field enhancements are computed and plotted in Figures 3(a) and (b), respectively, for different nanogap thicknesses. The nonlocal simulation results merge with the local model for nanogap sizes above $10\,nm$, where the screening effect ceases to exist. The deviation between these two models is larger in the few nanometer gap scale. This difference becomes even more apparent at the extremely thin one nanometer gap size.[45] Note that the electron spill-out effect is not considered in the current nonlocal model but is not expected to affect the current results, since the spill-out effect will become prominent only for extremely subwavelength nanogaps that are less than 1nm.[59] The presented results are generic and applicable to all gap-plasmon nanostructures, especially when perfect absorptivity is achieved for ultrathin nanogaps. Subsequently, we perform a time domain study based on the TTM to compute the maximum electron temperature at the resonance wavelengths presented in Figure 3(a). The results of the calculated maximum electron temperatures are demonstrated in Figure 3(c), where both local and nonlocal models are used. Interestingly, while the enhancement of electric field is predicted to be higher by using the local model with respect to the nonlocal calculations under the same gap size (Figure 3(b)), the difference between the temperature of hot carriers derived from local and nonlocal models is negligible (Figure 3(c)). Hence, the temperature variation trend in Figure 3(c) between local and non-local models is different from the trend observed for the field enhancement in Figure 3(b). The electromagnetic field enhancement in gap-plasmon metasurfaces is generally extremely localized in the nanogap region due to the plasmonic absorption resonance. However, thermal dissipation is naturally a highly diffusive process which requires more complicated plasmonic geometries to be spatially localized.[60] This minor difference is attributed to the very small nanostripe area ($40nm \times 40nm$) combined with the high thermal diffusivity of silver, which is equal to: $k_{diff} = k_L/(\rho \cdot C_p) = 0.17\ nm/fs$, where $k$ is the thermal conductivity, $\rho$ is the density, and $C_p$ is the specific heat capacity at constant pressure. The values of these silver parameters can be found in Table S1. Note that the total energy delivered to the metallic components of the nanostructure is the same for both local and nonlocal methods, as depicted in Figure S5, and follows similar trend to the electron temperatures shown in Figure 3(c).

On a relevant note, it has been correctly argued[44,61] that plasmonic nanostructures (such as the current nanocavities) can efficiently generate a much higher number of hot electrons compared to single plasmonic nanocrystals. Two options exist to compute the hot electron generation rate in our plasmonic system. The first one is the three-temperature model or modified TTM,[61,62] where the non-thermalized electron distribution is calculated in addition to the electron and lattice temperatures. The transient electronic excitation is incorporated in the TTM equations by using this technique, where the carrier density time dependency is computed by the relaxation-time approximation of the Boltzmann equation.[63,64] Instead of using this semi-classical approach, in our work we followed the more accurate quantum formulation of equation of motion based on the density matrix theory to calculate the generation rate of high energy non-thermalized electrons.[65] Hence, the generation rate of high energy (hot) electrons is calculated by using the formula:[39,65–67]

$$R_{HE-e^-} = \frac{1}{4}\frac{2}{\pi^2}\frac{e_0^2 E_F^2}{\hbar}\frac{(\hbar\omega - \Delta E_b)}{(\hbar\omega)^4}\int_S |E_{normal}|^2 ds, \qquad (8)$$

where, $e_0$ and $\hbar$ are electron charge and reduced Planck constant, respectively, $E_F$ is the Fermi energy of metal and $\Delta E_b$ is the energy barrier height that electrons need to overcome to be emitted from the surface. In general, plasmonic hot electrons are typically used for molecular surface chemistry or semiconductor applications.[39,65,68,69] Hence, $\Delta E_b$ is usually interpreted as the barrier height between metal and semiconductor[65] or between metal and molecules attached to its surface.[39] In our calculations, we choose $\Delta E_b = 1eV$ as a typical energy difference of molecules that can be attached to the metal-dielectric interface at the nanogap where the maximum field enhancement is present.[39,65]

The integral in Equation 8 consists of a surface integration of the normal electric field component ($E_{normal}$) along the surface of the metal at the nanogap. Figure 4(a) demonstrates the hot electron generation rate from the nanogap for increasing gap size thicknesses as a function of wavelength by using the local and nonlocal simulation models. The surface integration in Equation 8 is taken on the bottom surface of the nanostripe and the top surface of the silver substrate at the nanogap region where there is strong field enhancement. The hot electron generation rate reaches its maximum value exactly at the plasmonic resonance (compare Figures 4(a) and 2(a)). Again, the nonlocal model results are blueshifted but, interestingly, the hot electron generation rate is higher in this instance. The maximum value of the hot electron generation rate as a function of the nanogap thickness computed by the local and nonlocal models is shown in Figure 4(b). This result implies that the hot electron generation is substantially weakened for large gap thicknesses mainly due to the low field enhancement. However, the hot electron production rate rapidly increases for small nanogap sizes due to the localized and enhanced electric field. Interestingly, Figure 4(b) predicts higher rates of hot electron generation computed by the nonlocal model compared to the local model at small nanogap sizes. This phenomenon is not expected owing to the weaker electric field intensity in the nanogap dielectric region predicted by the nonlocal model (see Figures 2(c)-(d)). However, the strong electric field penetration into metals only in the nonlocal model makes the normal electric field component to have higher values along the metal interfaces compared to the local simulations. Nonlocal simulations are considered to be more realistic for extremely small nanogaps, since they match better the experimental data.[45,46] Hence, our findings prove for the first time that the value of the hot electron generation rate in extremely small nanogaps can be further increased when incorporating the quantum screening effect in metals compared to the usually performed classical local simulations. This result will be interesting for various applications that require the accurate computation of the hot electron generation from nanoscale regions.

To further understand the non-equilibrium temperature dynamics of the presented absorbers, we compute the electron and lattice temperatures derived from the TTM for various incident laser pulse durations from ultrashort femtosecond to a few picoseconds and nanoseconds. The relevant results are shown in Figures 5 and 6, where the gap thickness is fixed to $2nm$ and the remaining nanostripe dimensions are the same with before. Instead of fixing the laser power for different time regimes, which will result in high temperatures (above the lattice melting temperature) for longer pulse durations, we used different laser powers that result in the same steady state lattice temperature for all the time regimes involved in our studies. By following this approach, we achieve a fair comparison when demonstrating the various temperatures induced in different time regimes. As a result, we have adjusted the intensity of the laser pulse such that the maximum of lattice temperature always reaches the same value of $T_{L-max} = 500K$ in a long time duration, also known as steady-state, where electron and lattice temperatures merge and equilibrium is reached. More specifically, an ultrashort femtosecond laser has been used to obtain the results shown in Figure 5(a), where we observe a very sharp increase in the electron temperature mainly due to the ultrafast high-power laser illumination. The data in Figure 5 are always calculated in the point P shown in the inset of Figure 2(b). Figure 5(b) shows similar but more moderate behavior of electron temperature for $1ps$ incident laser pulse. The rapid increase in electron temperature occurs in the picosecond time scale regime in Figure 5(b)-(c). As it can be seen, not only the magnitude but also the time difference between the maximum values of the electron and lattice temperatures is decreased with respect to Figure 5(a). Hence, the non-equilibrium temperature dynamics are still present in the picosecond illumination case but are less pronounced compared to the ultrafast femtosecond excitation. At this stage electrons start to lose energy; however, electron carriers are not completely thermalized yet because there is still a large temperature difference between electrons and lattice. The hot carrier energy loss is attributed to electron-electron scattering at the initial stage and later on to electron-phonon interactions.**[61]** Note that the exact duration and intensity values of each incident pulse are given in the various panels of Figure 5.

As a next step, we excite the plasmonic absorber with slower nanosecond pulses and the induced electron and lattice temperature results are shown in Figure 5(d)-(f). The temperature of energetic electrons is substantially suppressed at the nanoscale temporal regime, as it is particularly evident in Figures 5(e) and 5(f). Electrons are slowly excited and have almost the same energy and temperature as the lattice in the case of nanosecond laser illumination. In addition, there is not evident temporal delay between the electron and lattice temperatures and the temperatures are mainly in equilibrium. Generally, hot/energetic electrons are always created in any plasmonic structure during strong light-matter interactions at the resonance regardless of the laser illumination pulse duration.**[61]** This relies on the fact that the linear momentum of electrons is not conserved along the surface of plasmonic nanostructures.**[44,65]** The formulation of the hot electron generation rate (Equation 8) has been derived based on this fact. Regarding the feasibility of experimental measurements, the main concern toward the detection of hot electrons is that they decay very fast on the order of femtoseconds.[23] Hence, the ratio of hot to thermal electrons (also known as Drude electrons or nearly thermalized electrons) is generally maximum at the femtosecond regime (clearly depicted in Figure 6 with more details provided in next paragraph) as long as the system (capable of generating intense electric field hot spots) is excited at the plasmonic resonance. The comparison of longer to shorter pulse responses shown in Figure 5 provides fruitful insights to experimentalists on how to directly detect hot carriers, which so far remains elusive or is a subject of intense debate.**[61]** It is of paramount importance to know the behavior of hot electrons for long and short pulses in the case of hot electron driven devices, such as detectors or photo catalyzers, where a large number of hot electrons is required to increase their sensitivity and improve their operation efficiency.[43]

We further study the difference between electron and lattice temperature dynamics by comparing the ratio of the maximum electron to lattice temperature, $\boldsymbol{T_{e-max}/T_{L-max}}$, achieved by various pulses as a function of different laser pulse durations. The results are depicted in Figure 6(a) and are calculated from the data presented before in Figure 5. Notable the electron temperature is substantially different compared to the lattice temperature for ultrafast femtosecond pulses. As the pulse duration ($\boldsymbol{\tau}$) becomes larger, this difference is significantly suppressed until almost constant values are reached at much slower nanosecond time scales. Another important phenomenon that can be derived by the results in Figure 5(a) is that the hot electron behavior rapidly vanishes after some picoseconds. This trend is analogous to the material changes due to the hot electrons ultrafast dynamics observed before[23] by measuring the extremely rapid variation in the reflection spectrum. The aforementioned method to estimate the hot electron dynamics is not an explicit way to accurately quantify the generation of hot electrons and is substantially different compared to our current work. It was observed that for thin gap-plasmon absorbers, the change in the reflection spectrum shows a sharp femtosecond peak followed by a relaxation tail of stable low reflectance at the resonance wavelength that eventually reaches steady state at the picosecond time scale regime. This implicit depiction of hot electrons transition to thermal electrons that affect the lattice temperature can also be derived by considering the time difference when the lattice temperature reaches its maximum value, $time_{T_{L-max}}$, with respect to the time when the maximum electron temperature, $time_{T_{e-max}}$, occurs. Hence, we plot in Figure 6(b) the ratio $time_{T_{L-max}}/time_{T_{e-max}}$ as a function of the incident pulse duration. The time difference between the electrons and lattice temperature maxima is significant for ultrashort laser pulses, exhibiting a substantial temporal delay due to the pronounced non-equilibrium temperature dynamics. When the pulse duration reaches the picosecond regime, this time delay is substantially suppressed. Finally, the ratio $time_{T_{L-max}}/time_{T_{e-max}}$ reaches unity values at the nanoscale time regime, meaning zero time delay between electron and lattice temperatures, which is the characteristic property of the thermally stable condition when temperature equilibrium is reached. The results presented in Figure 6 prove that the transition of hot electrons from energetic to thermalized carriers is ultrafast and mainly occurs in femtosecond to picosecond time scales.

It would be interesting to calculate the changes in the material properties leading to an ultrafast tunable absorption response of the nanostructure by using the current TTM-based simulations. The temperature can be substantially increased locally at the nanoscale by plasmonic structures. As a consequence, various material properties, such as permittivity, are expected to rapidly vary due to the increased electron and lattice temperatures, as has been shown in previous works.[70–72] Note that the permittivity plays a fundamental role to various light-matter interaction processes and, as a result, to the resonant absorption of the currently presented plasmonic metasurfaces. The temperature dependent permittivity of silver can be written by using the following Drude model:[47,73]

$$\varepsilon(\omega, T_e, T_l) = 1 - \frac{f_0 \omega_p^2}{\omega^2 - i\gamma(T_e,T_l)\omega} + \sum_j^n \frac{f_j \omega_p^2}{\omega_j^2 - \omega^2 + i\gamma_j(T_e,T_l)\omega}, \tag{9}$$

where $\omega$ is the incident laser frequency, $\omega_p$ is the plasma frequency, and $f_0$ is the dimensionless oscillator's strength. The summation in Equation 9 can be used to consider additional oscillators as correction terms in the Drude model. However, these higher order correction terms are usually weak, and we ignore them in our model for the sake of simplicity. The temperature dependence in the resulted permittivity mainly comes from the damping parameter $\gamma$, which is defined as the inverse of electron relaxation time $\tau_e$ and given by the formula:[47]

$$\gamma = \frac{1}{\tau_e} = A(T_e)^2 + B(T_L), \tag{10}$$

where $A$ and $B$ are the same constants used in Equation 3 that define the electron thermal conductivity of silver.[49] The parameters of silver used in Equations 9 and 10 are listed in Table S2 in the Supplementary Material. Note that the Drude model approach used before in the nonlocal (Equation 7) modeling results in the same permittivity at room temperature as Equation 9. The real and imaginary parts of silver permittivity derived by Equation 9 are plotted as a function of electron and lattice temperatures in Figure S6.

The temperature dependent real and imaginary parts of silver relative permittivity derived from the TTM under femtosecond laser illumination ($I_0 = 5\ GW/cm^2$) combined with Equations 9 and 10 for a $2nm$ thick gap metasurface design are shown in Figure S7. The electron temperature reaches its maximum value at the resonance of this nanostructure ($\sim 880\ nm$), leading to a considerable increase in the damping factor computed by Equation 10. This effect results in significant changes in the silver's complex permittivity values around the resonance wavelength, as demonstrated in Figure S7, while the room temperature (temperature-independent) permittivity remains unaltered. Interestingly, the silver permittivity is also spatially altered at high electron and lattice temperatures. This is shown in Figure S8 at the resonance wavelength ($880nm$) and exactly at the peak of the Gaussian pulse. The strong variation, especially in the imaginary part of the metal permittivity, will naturally lead to a rapid change in the absorption spectrum of the metasurface absorber, which is computed and demonstrated in Figure 7(a) in the case of a $2nm$ thick gap under femtosecond laser illumination ($I_0 = 5\ GW/cm^2$). The absorption properties of the presented plasmonic nanostructure become tunable since they are strongly affected by the induced high electron temperature in the femtosecond ultrafast time scale regime. This phenomenon of ultrafast extreme tunability in the absorption can be experimentally verified by pump-probe experiments based on transient absorption spectroscopy.[24] In addition, the ultrafast transient change in absorption can be useful in the emerging field of time-variant nanophotonics.[25–27] The computed quality-factor (Q-factor) and resonance wavelength of the ultrafast tunable absorption response are computed and plotted in Figures 7(b) and 7(c), respectively. The absorption resonance Q-factor is relatively low ~20, which is typical value for this type of plasmonic nanostructures,[1] but, interestingly, it can be tuned to even lower values combined with much lower absorption in the femtosecond time scale. The absorption resonance wavelength is also tunable in the same ultrafast regime, as depicted in Figure 7(c). Hence ultrafast all-optical switching of absorption can be obtained with the current plasmonic metasurface that can be used to extend the bandwidth of ultrathin plasmonic absorbers.[74] Note that this interesting effect ceases to exist for a $4nm$ gap thickness plasmonic absorber, as clearly depicted in Figure S9. The maximum electron temperature is decreased, as the nanogap thickness becomes larger, (see Figure 3(c)) and the effect of temperature in Equation 9 vanishes. At this point, it is noteworthy to mention that the nonlinear optical properties of some oxide materials, such as indium tin oxide (ITO), are determined by their free electron cloud ultrafast response generated when operate as epsilon-near-zero materials and excited by high input power lasers.**[75,76]** This effect can also be characterized by a temperature varying permittivity model and the current TTM-based simulations and will be subject of future work.

Finally, it is highly desirable during optical measurements of plasmonic nanostructures to know the ultimate limit in the laser power intensity that can be applied before leading to permanent sample damage, mainly due to heating. This metric can be especially useful for nonlinear optical experiments, such as second and third harmonic generation, where the conversion efficiency depends on and increases with the input laser intensity. Since the lattice temperature is directly dependent to the electron temperature due to the TTM theory, the effect of both these temperatures needs to be considered to find the ultimate limit of input laser intensity before the nanostructure is permanently damaged owing to lattice temperatures exceeding the melting point of the used materials. Figure 8 shows four different input laser pulse durations and their accompanying intensities, where the induced lattice temperature always reaches the melting temperature of silver $\boldsymbol{T_m} \sim \boldsymbol{1234.9\ K}$ (blue line) in all cases. Again, we used $\boldsymbol{2nm}$ gap thickness plasmonic absorbers with similar remaining nanostripe dimensions presented before. Based on these results, we can conclude that for ultrafast femtosecond laser pulses, the presented metasurface absorbers can sustain very high-power laser intensities on the order of tens $\boldsymbol{GW/cm^2}$. These high intensity values induce even larger electron temperatures (compared to Figure 2(b)) that can further change the transient absorption of the structure, as was shown before in Figure 7. On the contrary, in the nanosecond pulse duration case, the laser power intensity needs to be substantially reduced to few hundreds of $\boldsymbol{MW/cm^2}$ to not destroy the sample, which leads to much lower electron temperatures. It should be noted that we have used constant lattice density and heat capacity in these simulations for the sake of simplicity, alleviating the need to resort to more complicated molecular and quantum dynamic simulations. Temperature dependent physical properties that consider the nanoscale melting temperature rather than utilizing the bulk material properties should be used to compute the laser power damage thresholds more precisely. Using temperature dependent physical properties is also necessary in the case of ultrashort laser surface processing, where three phases (solid, fluid, and gas) are formed.[41,47] The current calculations are expected to be very interesting to nonlinear optical experiments based on the presented gap-plasmon plasmonic absorbers, since nonlinear efficiencies have been found to be significantly boosted by these structures[19,22] mainly due to the strong light-matter interaction in the nanogap[1] and can become even stronger when the laser input intensity is further increased.

# 4    Conclusions

To conclude, we studied the dynamics of hot electron relaxation processes in plasmonic nanocavities where strong light-matter interactions combined with quantum effects at the nanoscale are dominant. We demonstrated rapidly enhanced hot electron generation rates due to the boosted electric field inside the nanogap of these nanostructures, an interesting effect without a direct counterpart in bulk plasmonic materials. The performed comparison of longer to shorter pulse responses is expected to provide fruitful insights to experimentalists on how to directly detect hot carriers, currently a subject of intense debate,[61] which is of paramount importance to hot electron driven devices, such as detectors or photo catalyzers, where a large number of hot electrons is required to increase their sensitivity and improve their operation efficiency. It was shown that ultrashort femtosecond laser pulses result in high electron temperatures, where numerous hot electrons are produced. In the case of longer pulse durations, the hot electron behavior is rapidly suppressed due to both electron-electron collisions and electron-phonon interactions resulting to mainly thermalized electrons at the nanosecond time regime. The nanogap thickness is also crucial to achieve enhanced electron temperatures and hot electron generation. To correctly model extremely thin nanogaps, we introduced quantum electromagnetic simulations based on the nonlocal model to accurately compute the ultimate field enhancement in the plasmonic resonance and its effect on electron and lattice temperatures. The temperature dependent permittivity was also considered to study the ultrafast transient change in absorption leading to tunable plasmonic absorbers which can be useful in the emerging field of time-variant nanophotonics. Finally, the damage threshold of the proposed absorbers under various pulse illuminations was scrutinized. The presented results elucidate the role of hot electron generation and temperature dynamics in the response of gap-plasmon absorbers that can be used in a plethora of emerging applications, such as photocatalysis, photovoltaics, nonlinear optics, and photodetection.

## Supplementary Material

Supplementary Material is available online. It includes numerical modeling method details, materials parameter values, TTM verified by previously published works, sharp edge nanostripe results, oblique incidence illumination, energy loss delivered to the metallic components, real and imaginary parts of silver permittivity as function of electron and lattice temperatures, spatially varying real and imaginary parts of silver permittivity, and transient absorption contour plots.

## Acknowledgements

This work was partially supported by the NSF Nebraska Materials Research Science and Engineering Center (Grant No. DMR-1420645), Office of Naval Research Young Investigator Program (ONR-YIP) (Grant No. N00014-19-1-2384), National Science Foundation/EPSCoR RII Track-1: Emergent Quantum Materials and Technologies (EQUATE) under (Grant No. OIA-2044049), and the Jane Robertson Layman Fund from the University of Nebraska Foundation.

# Figures

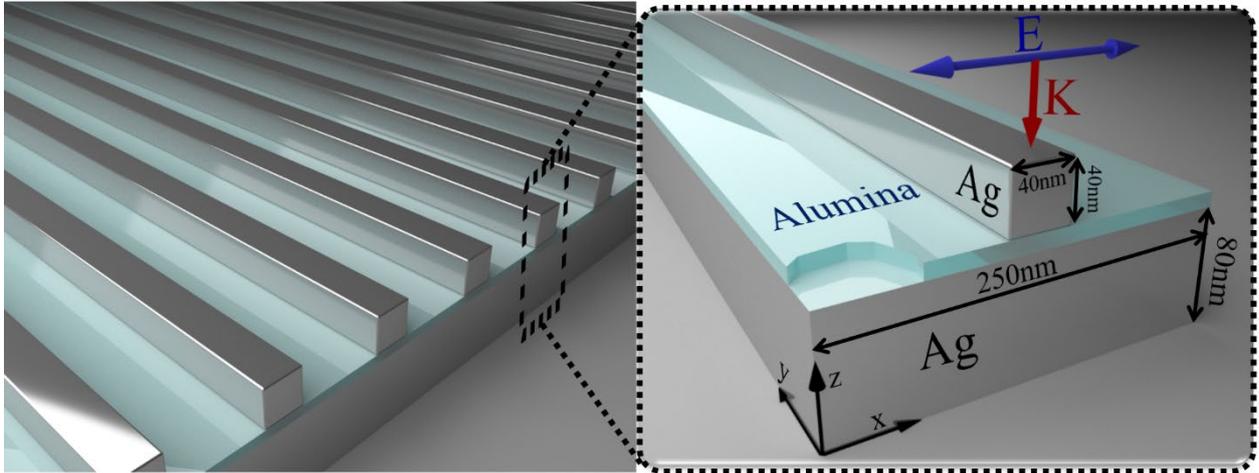

**Figure 1.** Geometry of the gap-plasmon metasurface absorber. Left panel: periodic array of silver nanostripes placed over a silver substrate and an ultrathin dielectric nanogap made of alumina. Inset right panel: unit cell of the gap-plasmon absorber showing the dimensions, materials, and polarization of the incident light.

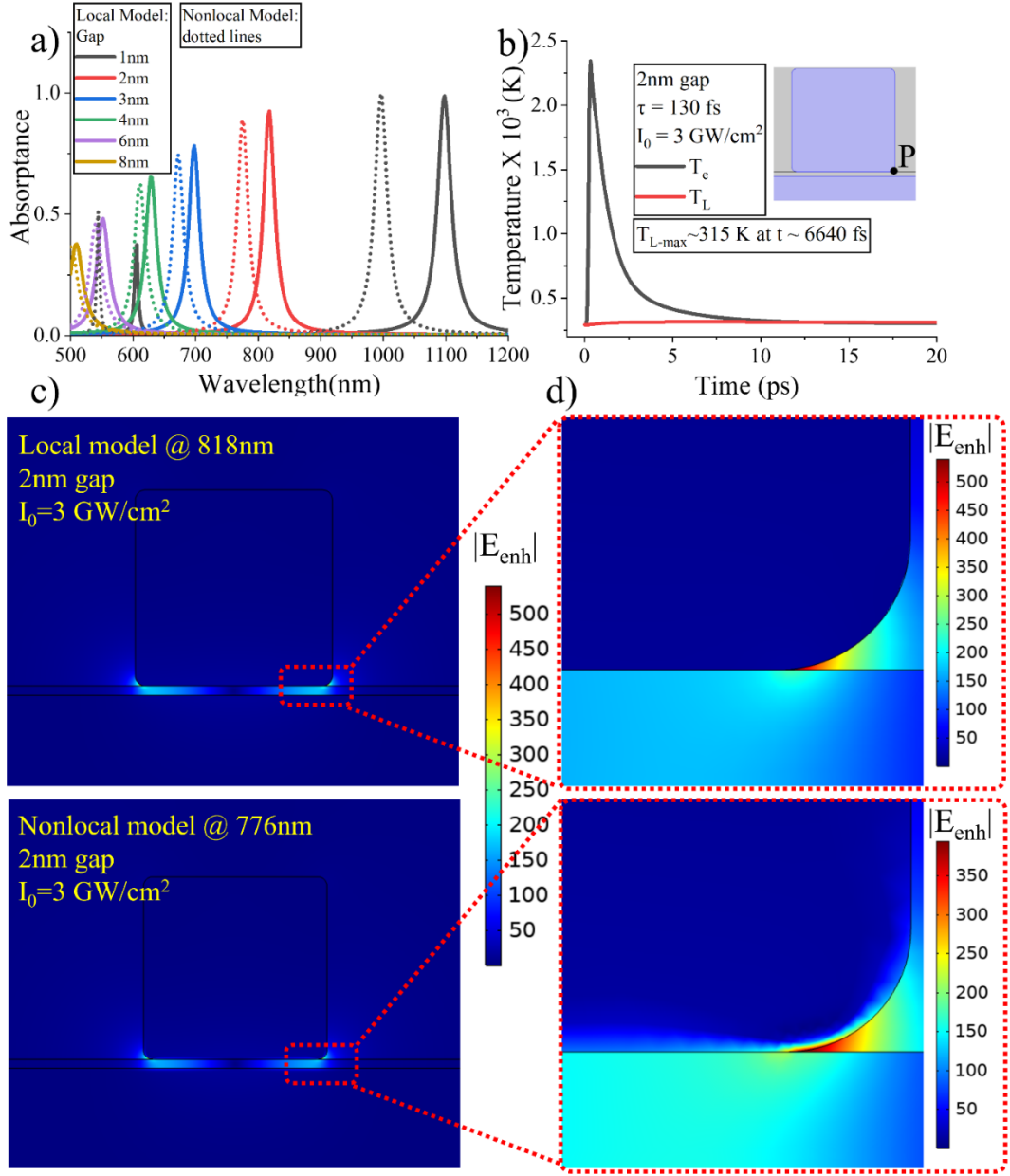

**Figure 2.** a) Resonant absorptance spectra for increasing nanogap thicknesses calculated by using the local (solid lines) and nonlocal (dotted lines) simulation models. b) Electron and lattice temperature dynamics in the case of $2nm$ thick nanogap induced by an ultrafast femtosecond laser. The initial room temperature is $T_0 = 293\ K$ and the local model is used in these non-equilibrium thermal simulations. Inset: geometry and point $P$ where the temperatures are computed. c) Electric field enhancement distribution, $E_{enh}$, for the same laser conditions used in (b) derived from local (top panel) and nonlocal (bottom panel) simulation models. The colorbar is the same in these two panels. d) Magnified regions close to the nanostripe corner in the results shown in (c) but now using different colorbar scales.

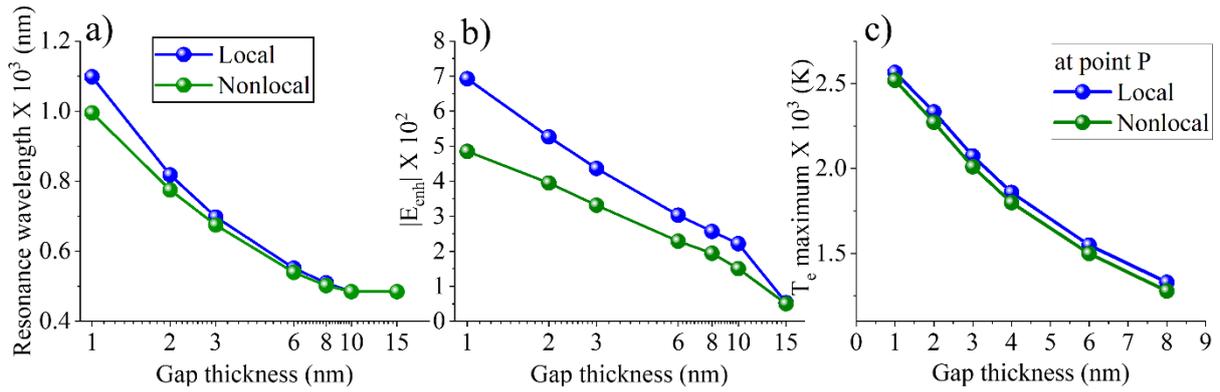

**Figure 3.** Local and nonlocal calculations of a) absorption resonance wavelengths (computed by Figure 2(a)), b) maximum electric field enhancements, and c) maximum electron temperatures. All these results are plotted versus the nanogap thickness. The induced maximum electron temperature values are calculated at point *P* depicted in the inset of Figure 2(b).

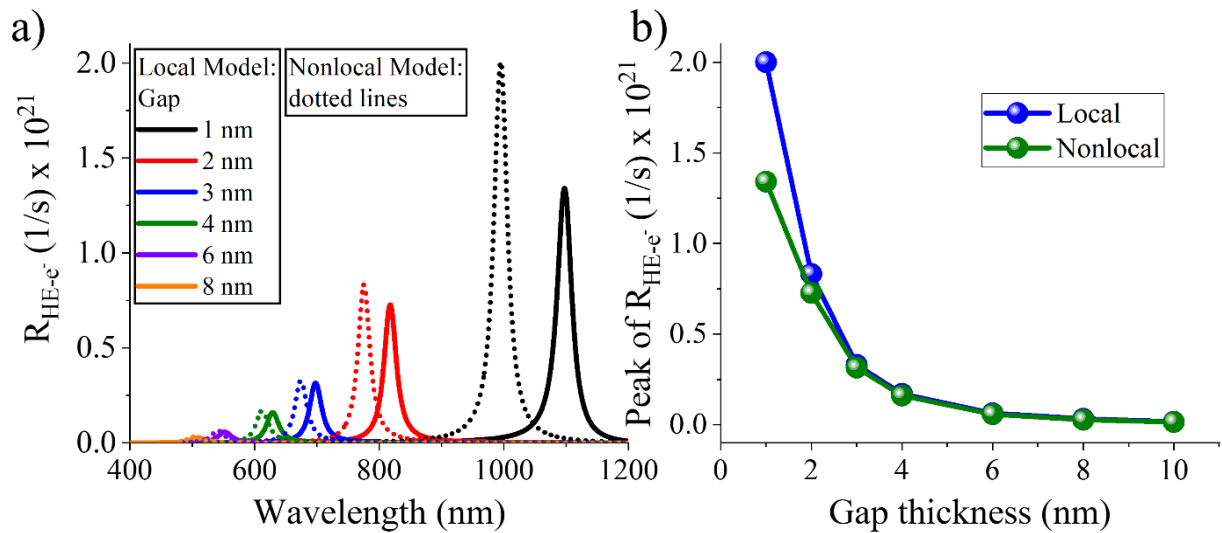

**Figure 4.** a) Generation rate of hot electrons as a function of the wavelength excitation for various nanogap thicknesses computed by the local and nonlocal model. b) The maximum hot electron generation rate at the resonance peaks shown in (a) calculated by local and nonlocal simulations.

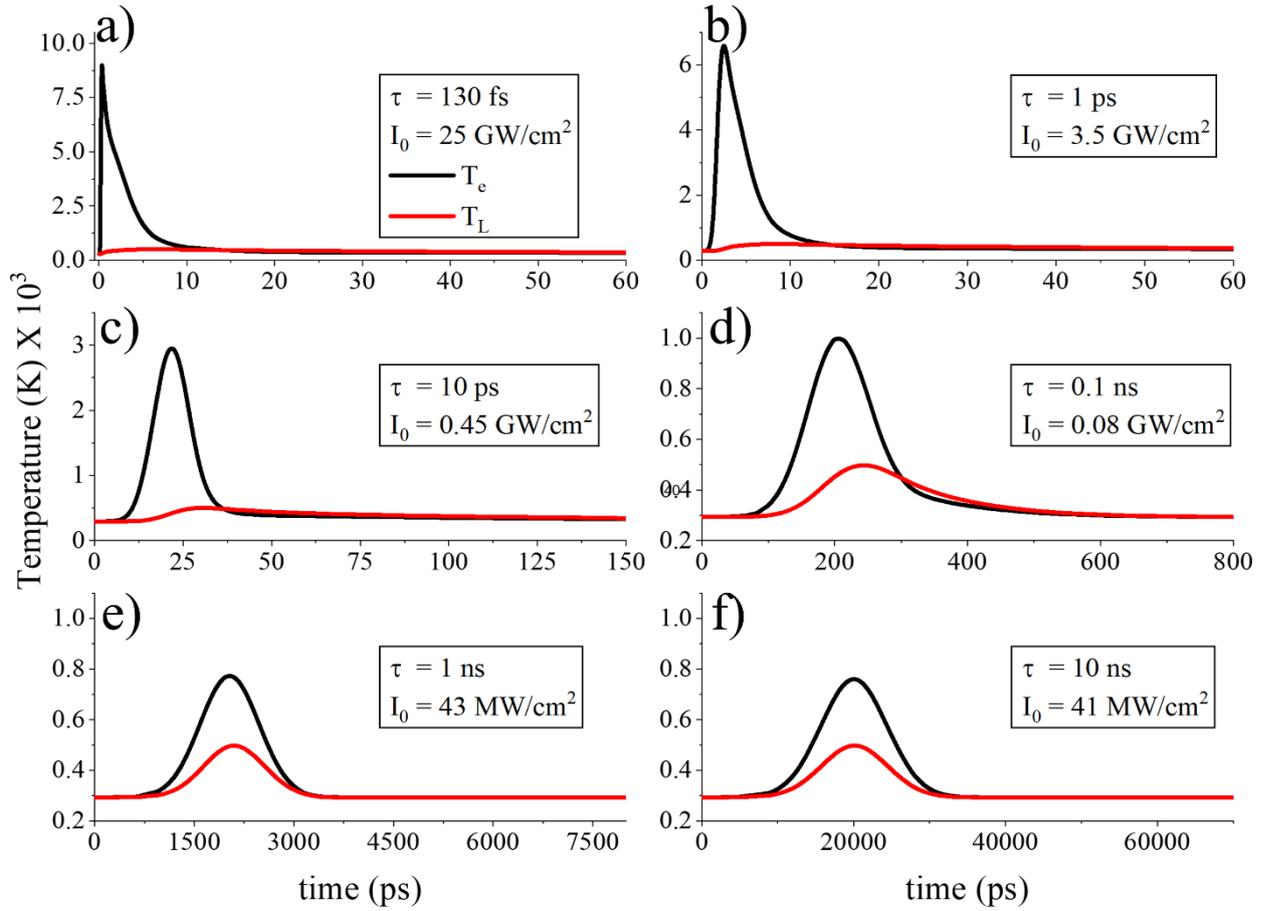

**Figure 5.** Electron (black lines) and lattice (red lines) temperature dynamics for (a), (b) ultrashort femtosecond, (c), (d) picosecond, and (e), (f) slow nanosecond incident pulse durations. All the data are calculated in the point P shown in the inset of Figure 2(b).

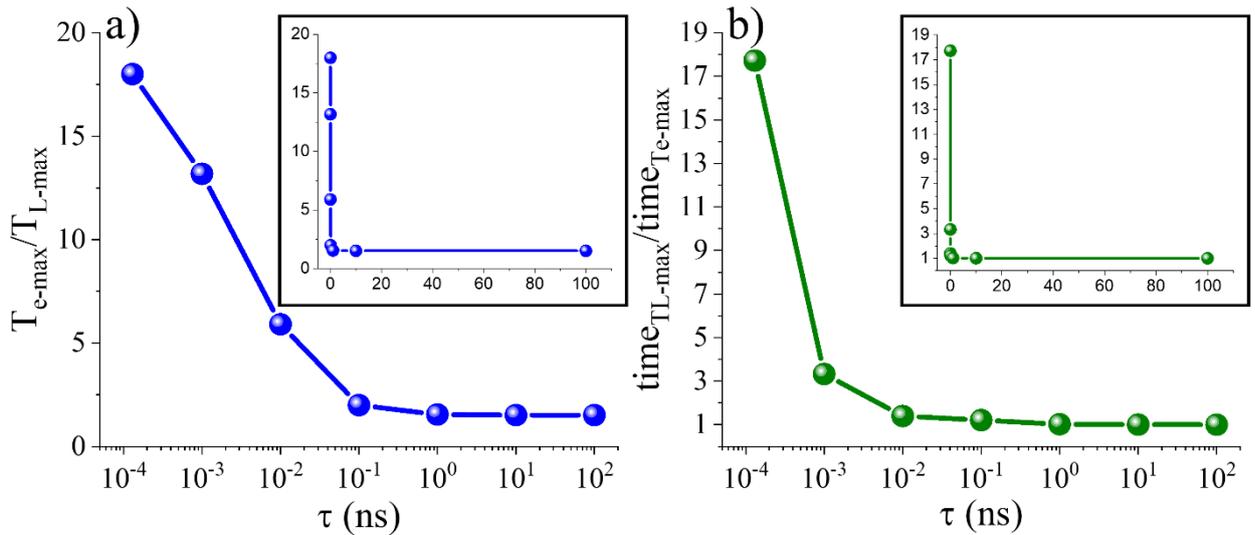

**Figure 6.** a) The ratio of the maximum electron over lattice temperature. b) The ratio of the time when $T_{e-max}$ occurs over the time when the lattice temperature reaches to $T_{L-max}$. Both ratios are plotted in logarithmic scale as a function of the incident pulse duration. These results are calculated from the data presented in Figure 5. The insets show the corresponding plots in linear scale. All the data are calculated in the point P shown in the inset of Figure 2(b).

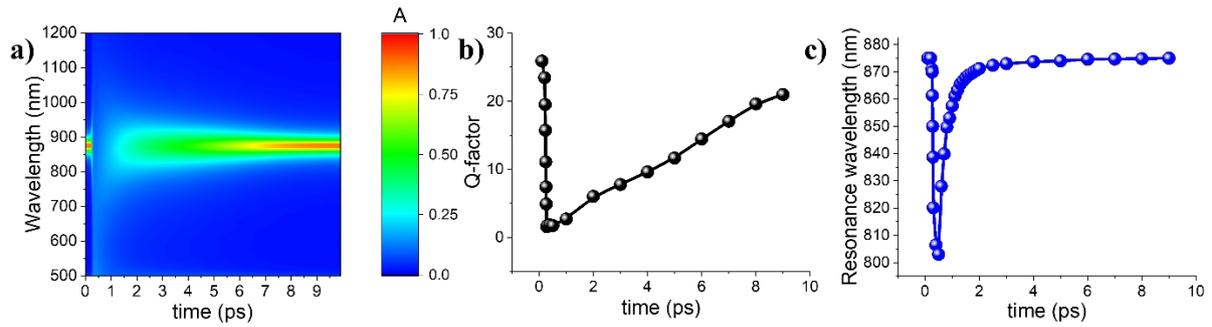

**Figure 7.** a) Tunable ultrafast transient absorption as a function of wavelength and time. b) The computed Q-factor of the absorption resonance response varying in time. c) The tunable absorption resonance wavelength is shifted as a function of time.

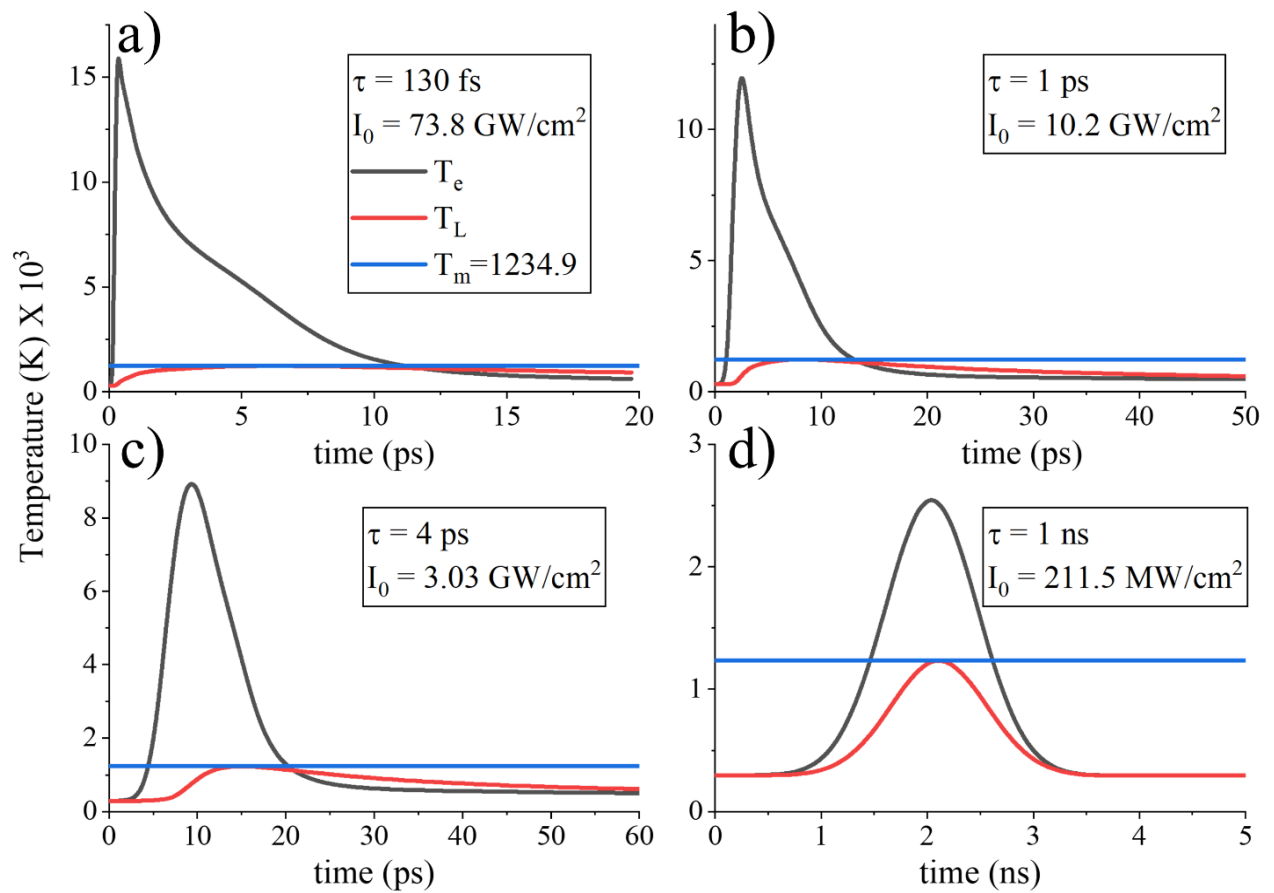

**Figure 8.** Damage (breakdown) study of the presented gap-plasmon metasurfaces. The induced lattice and electron temperatures for a) ultrashort femtosecond, b) one picosecond, c) four picoseconds, and d) slower one nanosecond laser pulse illuminations. The constant blue line depicts the melting temperature of silver.

21# References

[1] J.J. Baumberg, J. Aizpurua, M.H. Mikkelsen, D.R. Smith, Extreme nanophotonics from ultrathin metallic gaps, Nature Materials. 18 (2019) 668–678.

[2] G.M. Akselrod, J. Huang, T.B. Hoang, P.T. Bowen, L. Su, D.R. Smith, M.H. Mikkelsen, Metasurfaces: Large-Area Metasurface Perfect Absorbers from Visible to Near-Infrared (Adv. Mater. 48/2015), Advanced Materials. 27 (2015) 7897–7897.

[3] W. Li, Z.J. Coppens, L.V. Besteiro, W. Wang, A.O. Govorov, J. Valentine, Circularly polarized light detection with hot electrons in chiral plasmonic metamaterials, Nature Communications. 6 (2015) 8379.

[4] P. Yu, L.V. Besteiro, Y. Huang, J. Wu, L. Fu, H.H. Tan, C. Jagadish, G.P. Wiederrecht, A.O. Govorov, Z. Wang, Broadband Metamaterial Absorbers, Advanced Optical Materials. 7 (2019) 1800995.

[5] C. Argyropoulos, K.Q. Le, N. Mattiucci, G. D'Aguanno, A. Alù, Broadband absorbers and selective emitters based on plasmonic Brewster metasurfaces, Phys. Rev. B. 87 (2013) 205112.

[6] T. Guo, C. Argyropoulos, Tunable and broadband coherent perfect absorption by ultrathin black phosphorus metasurfaces, J. Opt. Soc. Am. B, JOSAB. 36 (2019) 2962–2971.

[7] Y. Cui, Y. He, Y. Jin, F. Ding, L. Yang, Y. Ye, S. Zhong, Y. Lin, S. He, Plasmonic and metamaterial structures as electromagnetic absorbers, Laser & Photonics Reviews. 8 (2014) 495–520.

[8] J. Hao, L. Zhou, M. Qiu, Nearly total absorption of light and heat generation by plasmonic metamaterials, Phys. Rev. B. 83 (2011) 165107.

[9] C. Argyropoulos, Electromagnetic Absorbers Based on Metamaterial and Plasmonic Devices, Forum for Electromagnetic Research Methods and Application Technologies (FERMAT). 1 (2014) 1–14.

[10] C. Argyropoulos, E. Kallos, Y. Zhao, Y. Hao, Manipulating the loss in electromagnetic cloaks for perfect wave absorption, Opt. Express, OE. 17 (2009) 8467–8475.

[11] S. Ray, M. Takafuji, H. Ihara, Amino-acid-based, lipid-directed,in situsynthesis and fabrication of gold nanoparticles on silica: a metamaterial framework with pronounced catalytic activity, Nanotechnology. 23 (2012) 495301.

[12] S. Mubeen, J. Lee, N. Singh, S. Krämer, G.D. Stucky, M. Moskovits, An autonomous photosynthetic device in which all charge carriers derive from surface plasmons, Nature Nanotechnology. 8 (2013) 247–251.

[13] M.L. Brongersma, N.J. Halas, P. Nordlander, Plasmon-induced hot carrier science and technology, Nature Nanotechnology. 10 (2015) 25–34.

[14] B. C. Marin, J. Ramírez, S. E. Root, E. Aklile, D. J. Lipomi, Metallic nanoislands on graphene: a metamaterial for chemical, mechanical, optical, and biological applications, Nanoscale Horizons. 2 (2017) 311–318.

[15] C. Kuppe, K.R. Rusimova, L. Ohnoutek, D. Slavov, V.K. Valev, "Hot" in Plasmonics: Temperature-Related Concepts and Applications of Metal Nanostructures, Advanced Optical Materials. 8 (2020) 1901166.

[16] W. Li, J. Valentine, Metamaterial Perfect Absorber Based Hot Electron Photodetection, Nano Lett. 14 (2014) 3510–3514.

[17] Y. Liu, Y. Chen, J. Li, T. Hung, J. Li, Study of energy absorption on solar cell using metamaterials, Solar Energy. 86 (2012) 1586–1599.

[18] J. Kong, A.H. Rose, C. Yang, X. Wu, J.M. Merlo, M.J. Burns, M.J. Naughton, K. Kempa, Hot electron plasmon-protected solar cell, Opt. Express, OE. 23 (2015) A1087–A1095.

[19] C. Ciracì, M. Scalora, D.R. Smith, Third-harmonic generation in the presence of classical nonlocal effects in gap-plasmon nanostructures, Phys. Rev. B. 91 (2015) 205403.

[20] B. Jin, C. Argyropoulos, Enhanced four-wave mixing with nonlinear plasmonic metasurfaces, Scientific Reports. 6 (2016) 28746.

**Supplementary Material**

**Unraveling the temperature dynamics and hot electron generation in tunable gap-plasmon metasurface absorbers**

*Larousse Khosravi Khorashad and Christos Argyropoulos\**

Department of Electrical and Computer Engineering, University of Nebraska-Lincoln, Lincoln, NE, 68588, USA
*E-mail: christos.argyropoulos@unl.edu

**Numerical Modeling Method Details**

In this work, we used COMSOL Multiphysics for all the performed numerical calculations. COMSOL Multiphysics is a commercial software based on the finite element method. The Radio Frequency (RF) module was used to calculate the plasmonic resonances along with the electromagnetic heating computed by Equation 6 and generation rate of high energy (hot) electrons given by Equation 8 in the main text. The entire geometry was excited with an electromagnetic wave via a port excitation propagating from the top to the bottom. Periodic boundary conditions were placed on both geometry sides located at a distance equal to the structure's periodicity. Furthermore, we used two separate Heat Transfer (HT) COMSOL modules to solve the TTM coupled partial differential equations (Equations 1 and 2 in main text) and, subsequently, calculate the electron and lattice temperatures in time domain. Note that Equations 1 and 2 do not have the conventional heat equation form. The additional terms, such as the electron-lattice coupling factor and the source term in Equation 1, are introduced as supplementary heat sources in the first HT module dedicated to solving the electron temperature. Similarly, the coupling factor in Equation 2 is introduced as a heat source in the second HT module, which is used to compute the lattice temperature. The two HT modules are coupled by the electron and lattice temperature variables. The computed electromagnetic heating derived by Equation 6 using the RF module was coupled to the HT module via the heat source definition.

Periodic conditions were chosen for the metallic parts and the entire surrounding was kept at room temperature as the initial temperature condition. The heat source was applied to the metallic parts since only them possess an imaginary part of permittivity. All parts of the geometry in RF and HT modules are coupled via the Multiphysics Electromagnetic Heating (emh) module.

The electric current density $\vec{j}(\vec{r})$ in the nonlocal model of Equation 7 is computed by using the weak form partial differential equation COMSOL module. Then, the calculated current density $\vec{j}(\vec{r})$ is introduced as a weak contribution to the COMSOL frequency domain electromagnetic solver. This weak contribution is added as an additional polarization term in the electromagnetic wave equation that now becomes: $\nabla \times \nabla \times \vec{E}(\vec{r}) - k_0^2 \vec{E}(\vec{r}) = i\omega\mu_0 \vec{j}(\vec{r})$, which is solved by the COMSOL frequency domain electromagnetic solver. Finally, the continuity in the normal component of the electric displacement field between silver and the dielectric nanogap needs to be manually introduced in the COMSOL weak form module as an additional Dirichlet boundary condition. This is due to the presence of spatial derivatives in the nonlocal current density equation given by Equation 7.

# Figures and Tables

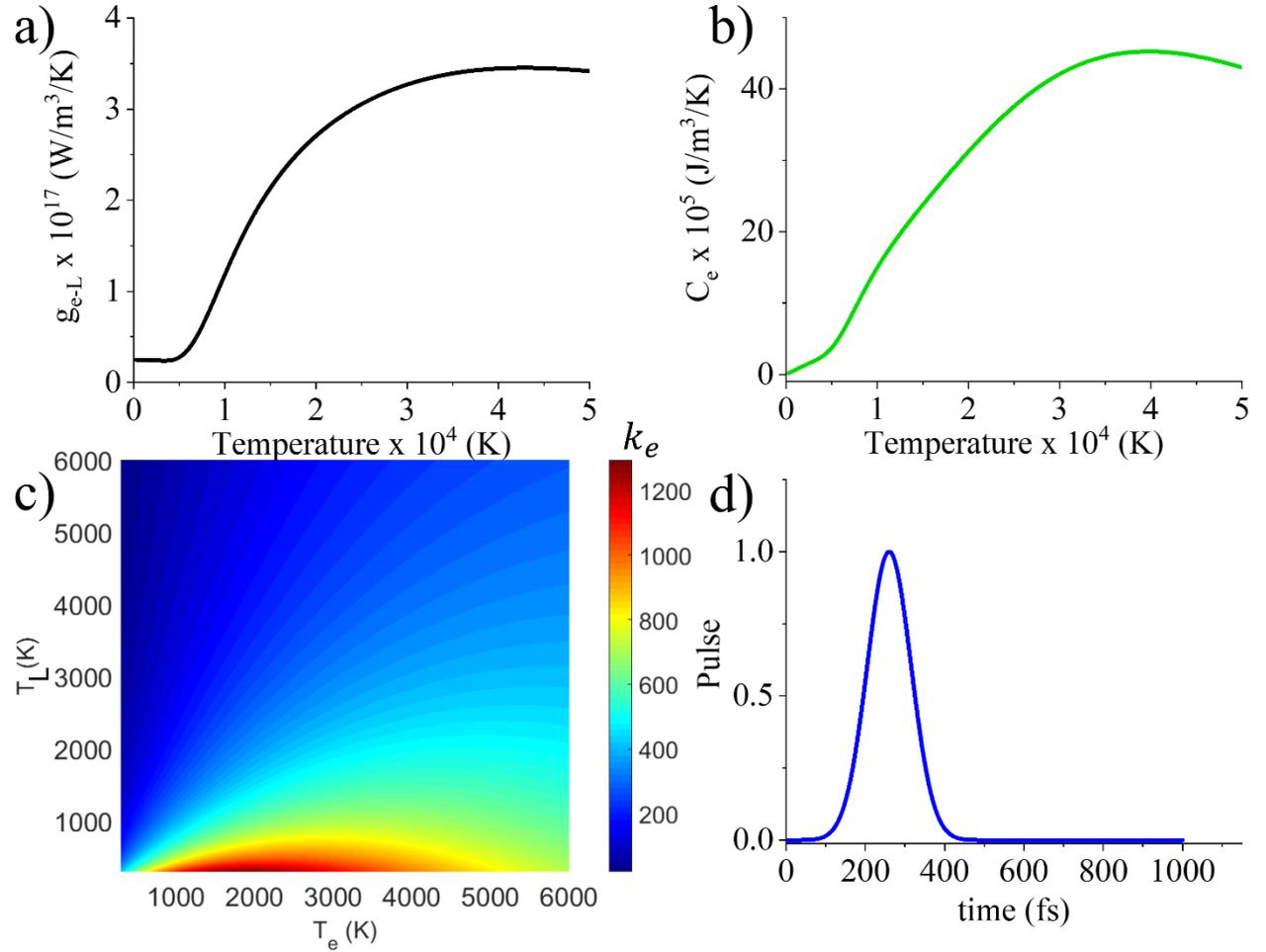

**Figure S1.** a) Electron-lattice coupling factor, b) electron heat capacity, and c) electron thermal conductivity derived from Equation (3). The material characterized by these parameters is silver. d) The envelope of the femtosecond Gaussian pulse with maximum amplitude at $t = 3\tau$, where $\tau = 130 fs$.

Table S1. Various parameters used in our simulations.

| Parameter | Value | Unit |
|---|---|---|
| Density ($\rho$) of silver | 10.49 | g/m$^3$ |
| Density ($\rho$) of alumina | 3.95 | g/m$^3$ |
| Density ($\rho$) of air | 1205 | g/m$^3$ |
| Lattice thermal conductivity ($k_L$) of silver | 428[W/(m*K)] | W/m/K |
| Thermal conductivity ($k$) of alumina | 18[W/(m*K)] | W/m/K |
| Thermal conductivity ($k$) of air | 0.024 | W/m/K |
| Specific heat capacity ($C_p$) of silver (lattice) | 0.233 | J/kg/K |
| Specific heat capacity ($C_p$) of alumina | 880 | J/kg/K |
| Specific heat capacity ($C_p$) of air | 1005 | J/kg/K |
| Permittivity of silver | Ref.[1] | - |
| Permittivity of alumina | 3.13 | - |
| Permittivity of air | 1 | - |

**Table S2.** Silver parameters used in Equation 3, 9 and 10 taken from Refs.[2,3]

| Parameter | Value | Unit |
|---|---|---|
| $\omega_p$ | 9.01 | eV |
| $f_0$ | 0.845 | - |
| $A$ | $0.932 \times 10^7$ | 1/s/K$^2$ |
| $B$ | $1.02 \times 10^{11}$ | 1/s/K |

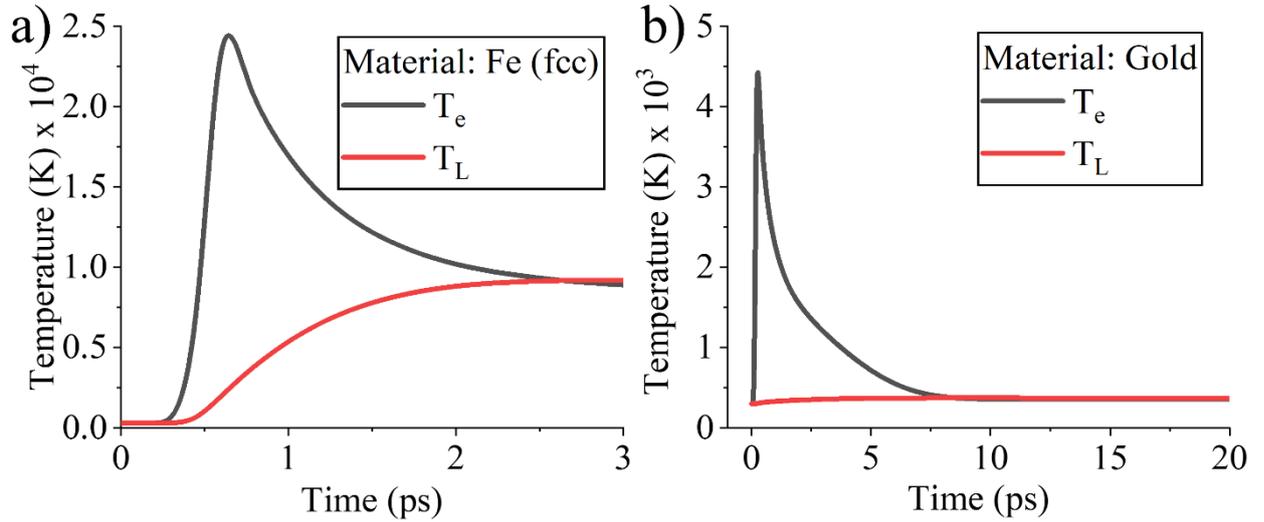

**Figure S2.** Electron and lattice temperatures computed by the current multiphysics simulations and obtained at the surface of a) bulk iron (fcc) and b) 200nm thick flat gold film. The results perfectly agree with previous published works based on analytical approaches.[2,4] All the parameters used in these simulations are identical to the previous analytical works.[2,4]

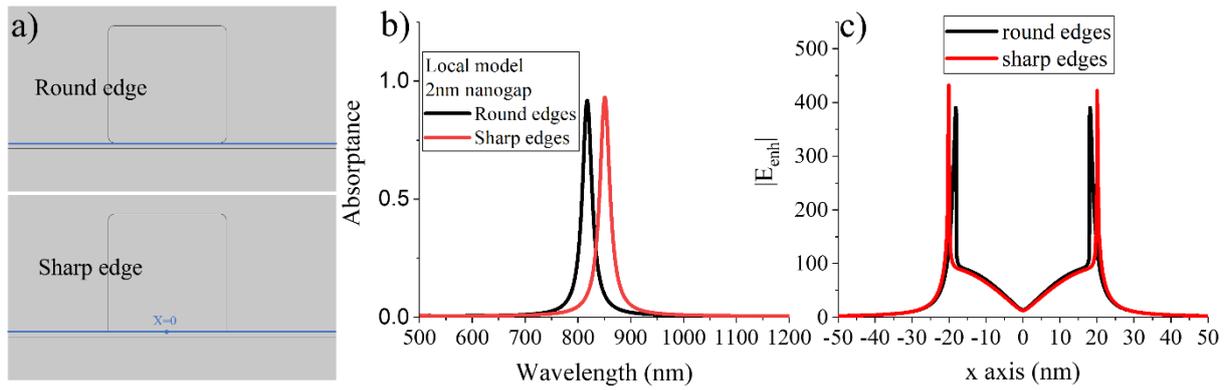

**Figure S3:** a) Round (top panel) and sharp (bottom panel) edge nanostripe geometries in the case of a 2nm nanogap metasurface absorber. The round scenario is the same with the one used in the main paper. b) Absorptance spectrum of the two geometries demonstrated in (a). c) Electric field enhancement computed along the blue line in (a) (bottom of the nanostripe) at each corresponding absorptance resonance: 818nm (round edge) and 852nm (sharp edge).

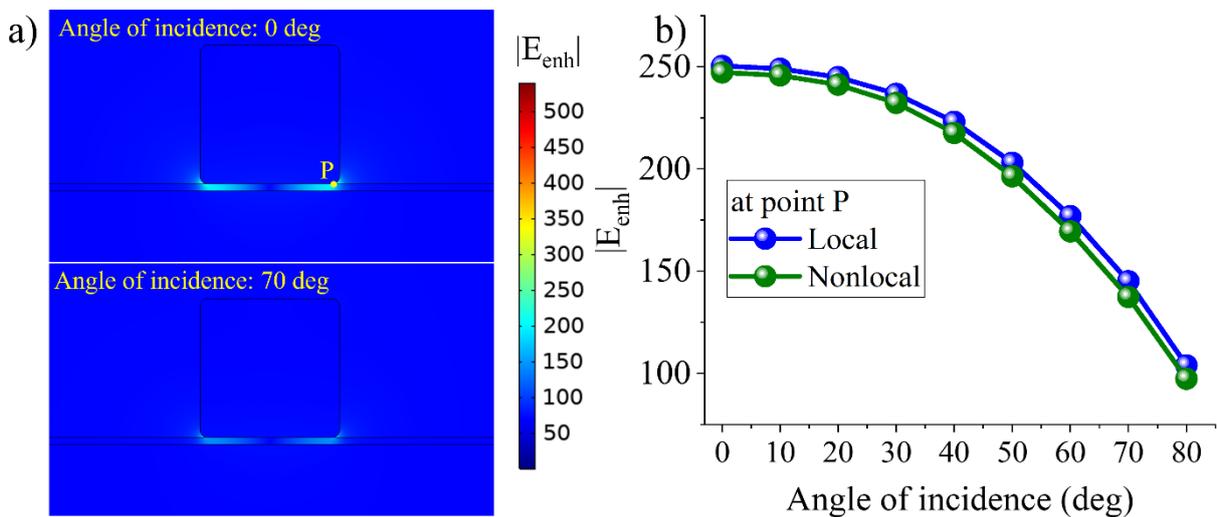

**Figure S4:** a) Computed electric field enhancement distribution by using the local model for normal (top panel) and 70° oblique (bottom panel) incident illumination. b) Electric field enhancement as a function of incidence angle for local and nonlocal models computed at point P shown in caption (a). These calculations are for the 2nm nanogap metasurface.

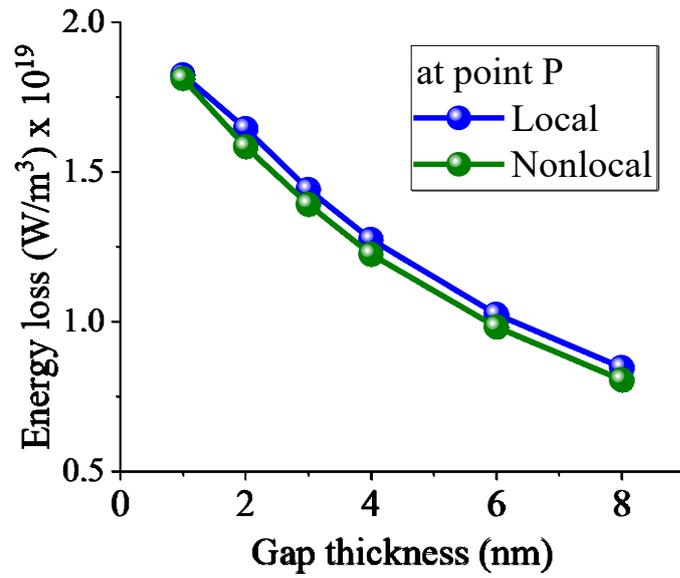

**Figure S5**: Average energy loss delivered to the metallic components of the nanostructure computed at point P shown in Figure 3(c) for a 2nm nanogap metasurface illuminated by a laser intensity of 3GW/cm$^2$ in the case of local and nonlocal models.

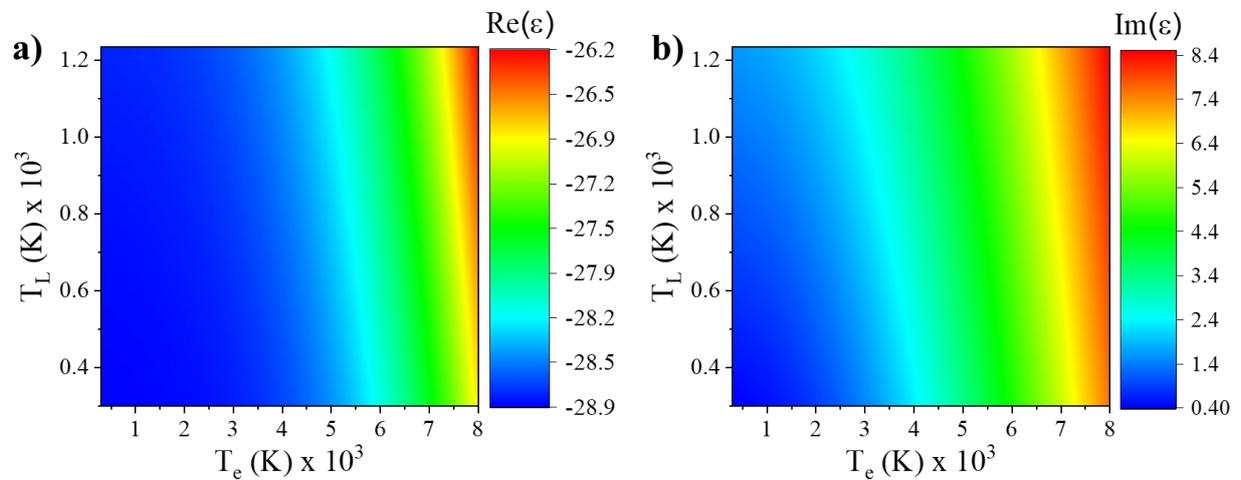

**Figure S6**: (a) Real and (b) imaginary parts of silver permittivity plotted as a function of electron (x-axis) and lattice (y-axis) temperatures. The results are computed under femtosecond laser illumination at the resonance wavelength (818nm) of a 2nm nanogap metasurface.

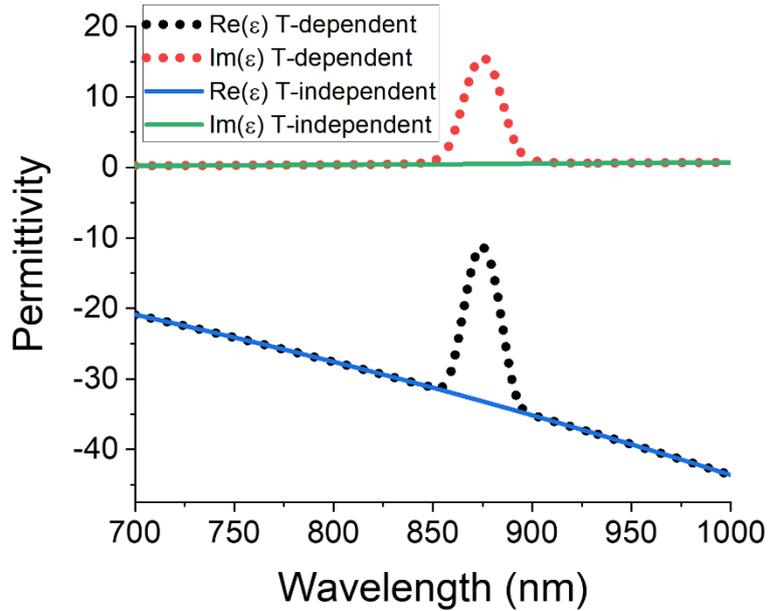

**Figure S7.** The local temperature dependent real and imaginary parts (dashed lines) of silver relative permittivity at point P (see inset of Figure 2(b)) derived from the TTM under femtosecond laser illumination ($I_0 = 5 GW/cm^2$) combined with Equations 9 and 10 in the main paper for a $2nm$ thick gap metasurface design. The silver permittivity at room temperature is also depicted with the solid lines. The electron temperature reaches its maximum value at the resonance of this nanostructure (~$880\ nm$), leading to a considerable increase in the damping factor computed by Equation 10 in the main paper. This effect results in significant changes in the silver's complex permittivity values around the resonance wavelength, as demonstrated in Figure S7, while the room temperature (temperature-independent) permittivity remains unaltered.

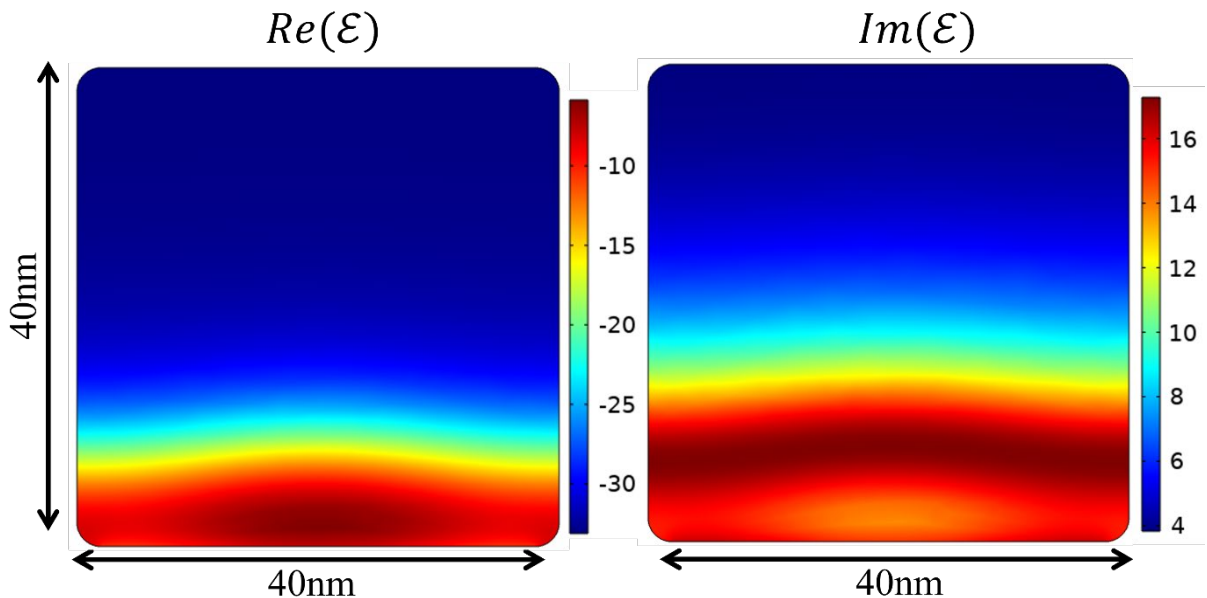

**Figure S8.** Spatially varying real and imaginary parts of silver permittivity plotted in a cross section of the silver nanostripe under femtosecond laser illumination ($I_0 = 5\ GW/cm^2$). These permittivity values are obtained at the resonance wavelength of $880nm$ and exactly at the Gaussian pulse peak. The silver substrate does not have a pronounced permittivity change (not shown here) because is mainly used as the reflector in the formed nanocavity and has much larger dimensions compared to the nanostripe.

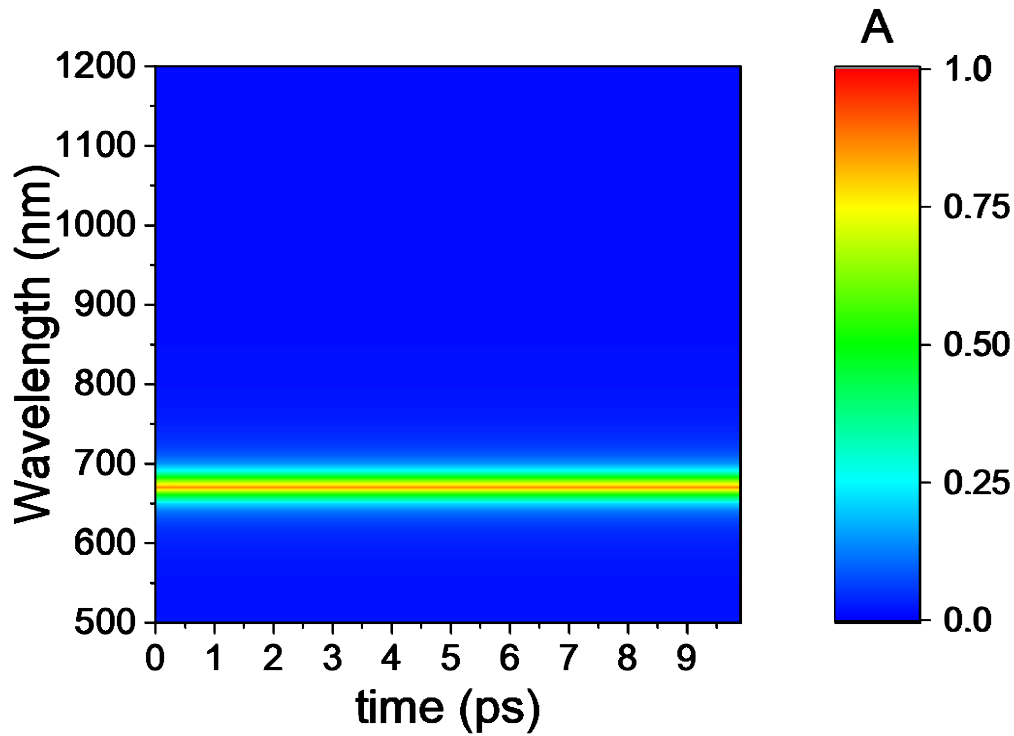

**Figure S9.** Computed absorptance distribution versus time and wavelength obtained by using the temperature varying silver permittivity values defined by Equation 9 in the main paper for 4nm gap thickness plasmonic absorber. Femtosecond ultrafast laser illumination is used with laser input intensity equal to $I_0 = 5\ GW/cm^2$. No temporal change in the absorption spectrum is obtained on vast contrast to Figure 7 in the main paper (2nm gap plasmonic absorber case).